\documentclass[%
final,      
paper=a4, paper=portrait, pagesize=auto, 
fontsize=9pt, 
american, 
bibliography=totoc,
twocolumn,
lineno
]{scrartcl} 

\usepackage[utf8]{inputenc}
\usepackage[extendedchars, encoding, multidot, space,filenameencoding=utf8]{grffile}
\usepackage{dsfont}
\usepackage{textcomp}
\usepackage{physics}
\usepackage{amsfonts}
\usepackage{amsmath,amssymb}

\usepackage{graphicx}
\usepackage{caption}
\usepackage{epsfig}

\usepackage{color}
\usepackage[normalem]{ulem}  

\usepackage{titlesec}
\titleformat{\section}
  {\normalfont\rmfamily\Large\bfseries}
  {\thesection}{1em}{}

\titleformat{\subsection}
  {\normalfont\rmfamily\large\bfseries}
  {\thesection}{1em}{}
\titleformat{\chapter}
  {\normalfont\rmfamily\huge\bfseries}
  {\thesection}{1em}{}
  \titleformat{\chaptertitlename}
  {\normalfont\rmfamily\huge\bfseries}
  {\thesection}{1em}{}

\usepackage[super, comma, square, sort&compress]{natbib}
\usepackage{placeins}

\usepackage{xargs}  
\usepackage[pdftex,dvipsnames]{xcolor}  
\usepackage{physics} 
\usepackage{siunitx} 
\usepackage{hyperref} 
\usepackage{listings} 

\definecolor{clr-background}{HTML}{1E1E1E}
\definecolor{clr-text}{HTML}{C0C0C0}
\definecolor{clr-string}{HTML}{E5C365}
\definecolor{clr-namespace}{RGB}{0,0,0}
\definecolor{clr-preprocessor}{RGB}{128,128,128}
\definecolor{clr-keyword}{HTML}{8DCBE2}
\definecolor{clr-type}{HTML}{8DCBE2}
\definecolor{clr-variable}{RGB}{255,255,255}
\definecolor{clr-constant}{RGB}{111,0,138} 
\definecolor{clr-comment}{HTML}{7F9F7F}

\lstdefinestyle{VS2018}{
    language=[Visual]C++,
    basicstyle=\color{clr-text}\footnotesize\ttfamily,
    numbers=left,
    numberstyle=\tiny,
    numbersep=5pt,
    extendedchars=true,
    frame=b,
    stringstyle=\color{clr-string}\ttfamily,
    xleftmargin=17pt,
    framexleftmargin=17pt,
    framexrightmargin=5pt,
    framexbottommargin=4pt,
    commentstyle=\color{clr-comment},
    morecomment=[l]{//}, 
    morecomment=[s]{/*}{*/}, 
    showstringspaces=false,
    morekeywords={ abstract, event, new, struct,
    as, explicit, null, switch,
    base, extern, object, this,
    bool, false, operator, throw,
    break, finally, out, true,
    byte, fixed, override, try,
    case, float, params, typeof,
    catch, for, private, uint,
    char, foreach, protected, ulong,
    checked, goto, public, unchecked,
    class, if, readonly, unsafe,
    const, implicit, ref, ushort,
    continue, in, return, using,
    decimal, int, sbyte, virtual,
    default, interface, sealed, volatile,
    delegate, internal, short, void,
    do, is, sizeof, while,
    double, lock, stackalloc,
    else, long, static,
    enum, namespace, string},
    keywordstyle=\color{clr-type},
    identifierstyle=\color{clr-variable},
    backgroundcolor=\color{clr-background},
    directivestyle=\color{clr-preprocessor}, 
}

\DeclareCaptionFont{white}{\color{white}}
\DeclareCaptionFormat{listing}{\colorbox{blue}{\parbox{\textwidth}{\hspace{15pt}#1#2#3}}}
\usepackage{pdfpages}
\usepackage{booktabs}

\usepackage[left=2.00cm,right=2.00cm,top=2.00cm,bottom=3.50cm]{geometry}
\usepackage[affil-it]{authblk}
\usepackage{abstract}

\begin{document}

\title{\normalfont\rmfamily\huge\bfseries Hydrodynamic Inflation of Ring Polymers under Shear}
\author[1]{M. Liebetreu \thanks{maximilian.liebetreu@univie.ac.at}}
\author[1]{C. N. Likos \thanks{christos.likos@univie.ac.at}}
\affil[1]{Faculty of Physics, University of Vienna, Boltzmanngasse 5, A-1090 Vienna, Austria}

\date{\vspace{-5ex}}


\twocolumn[
  \begin{@twocolumnfalse}
    \maketitle
    \begin{abstract}
      Hydrodynamic interactions as modeled by Multi-Particle Collision Dynamics can dramatically influence the dynamics of fully flexible, ring-shaped polymers 
        in ways not known for any other polymer architecture or topology.  We show
        that steady shear leads to an inflation scenario exclusive to ring polymers, which depends not only on Weissenberg number but also on contour length of the ring.
        By analyzing velocity fields of the solvent around the polymer, we show the existence of a hydrodynamic pocket
        which allows the polymer to self-stabilize at a certain alignment angle to the flow axis.
        This self-induced stabilization is accompanied by transitioning of the ring to a non-Brownian particle and a cessation of tumbling. 
        The ring swells significantly in the vorticity direction, and the horseshoe regions on the stretched and swollen ring 
        are effectively locked in place relative to the ring's center-of-mass.
        The observed effect is exclusive to ring polymers and stems from an interplay between hydrodynamic interactions and topology.
        Furthermore, knots tied onto such rings can serve as additional "stabilization anchors". 
        Under strong shear, the knotted section is pulled tight and remains well-localized while tank-treading from one horseshoe region to the opposite one in sudden bursts.
        We find knotted polymers of high contour length behave very similarly to unknotted rings of the same contour length, but small knotted rings feature a host of different configurations. 
        We propose a filtering technique for rings and chains based on our observations and suggest that strong shear could be used to tighten knots 
        on rings.
        \bigskip\bigskip
    \end{abstract}
  \end{@twocolumnfalse}
] 

\saythanks
\bigskip\noindent


\section*{Introduction}

    The influence of topology on the physical properties of macromolecules is profound and manyfold \cite{grosberg:aps:1993}. 
    The most striking manifestation of this fact is the vast richness of novel phenomena emerging in the behavior of ring polymers,
    both at the single-molecule level as well as for concentrated solutions and melts \cite{frank-nature-1975,stasiak-nature2-1996,micheletti-physrep-2011,kapnistos-natmat-2008,vlassopoulos:prl:2019}.  
    Of immediate interest are, for example, the interplay between molecular architecture, deformation and dynamics under shear, but also 
    how these affect stress distribution, viscosity, rheology and thixotropic behavior of sheared concentrated solutions. 
    The conformations of single molecules
    and the ensuing effective interactions \cite{grosberg-pnas-2004,grosberg-pre-2005,heermann-jcp-2010,narros-sm-2010}, 
    the self-organization of concentrated ring polymer solutions and melts \cite{vettorel:pb:2009,halverson1:jcp:2011,sakaue-prl-2011,grosberg:sm:2014,smrek:rpp:2014,smrek:jpcm:2015,smrek:acsml:2016}, 
    the viscosity and stress relaxation under shear \cite{kapnistos-natmat-2008,halverson:prl:2012,vlassopoulos:acsml:2013,vlassopoulos:mm:2015}
    as well as the possibility of emergence of novel, topologically glassy states for ring polymer melts 
    \cite{michieletto:pnas:2016,michieletto:polymers:2017,michieletto:prl:2017} are a few characteristic examples
    of the variety of properties unique to cyclic polymers. Concomitantly, the out-of-equilibrium behavior of rings under tailored
    microfluidic flows has also attracted considerable attention in the recent past \cite{li-mm-2015,hsiao-mm-2016,liebetreu-acsml-2018,sing:pre:2019}.
    
    As the probability to obtain unknotted polymers decreases exponentially with their contour length \cite{koniaris-prl-1991},
    the importance of knots on polymer behavior is an additional topic of intensive research activity. 
    Knots are naturally found on long DNA strands \cite{wasserman-science-1986,rybenkov-pnas-1993,marenduzzo-bj-2008}
    and proteins \cite{virnau-plos-2006,sulkowska-pnas-2012,tumanski-pnas-2017}
    while at the same time they can also be artificially \cite{arai-nature-1999,bao-prl-2003}
    or synthetically manufactured \cite{leigh-ac-2017}. The influence of such knots in equilibrium and relaxation properties 
    in the bulk \cite{grosberg-pre-1996,grosberg-prl-2000,katritch-pre-2000,grosberg-prl-2007,douglas-rings-JCP1-2010,douglas-rings-jcp2-2010,tubiana-mm-2013,narros-mm-2013,doyle-mm-2016}, 
    under confinement \cite{quake-prl-1994,lai-pre-2002,sheng-pre-2001,micheletti-mm-2012,doyle-acsml-2012,sakaue-sm-2013,dadamo-mm-2015,doyle-mm-2015,klotz-mm-2017,marenda-sm-2017}
    and under tension \cite{saitta-nature-1999,sheng-pre-2000,farago-epl-2002,doyle:pnas:2011,matthews-acsml-2012,poier-mm-2014,caraglio-prl-2015,doyle:prl:2018}
    has been thoroughly investigated. Recently, the sedimentation behavior of flexible, non-Brownian knots has been added to the host of counterintuitive phenomena \cite{gruziel-prl-2018}. 

    All polymer architectures (linear, star, dendritic, cross-linked and ring) are known to undergo tumbling under steady shear at 
    sufficiently high 
    shear rates \cite{ripoll-prl-2006,ripoll-epje-2007,huang-mm-2010,huang-epl-2011,fedosov-sm-2012,nikoubashman-mm-2010,jaramillo:jpcb:2018,liebetreu-acsml-2018,formanek:mm:2019}. 
    Moreover, it is known that linear chains never fully stretch under shear and they reach
    a stretched configuration only under so-called planar mixed flows that represent a combination of simple shear 
    and planar extension \cite{hur:pre:2002,woo:jcp:2003,sing:pre:2019}.
    Under simple shear, ring polymers show two dynamical modes: Tumbling (TB) and tank-treading (TT) \cite{frey-pre-2014}.
    During TB, the polymer's shape changes rapidly as the ring collapses and expands again, with beads at the horseshoe regions swapping to the opposite side.
    The transition between initial and final state can best be described by the flipping of a disc, except it collapses onto itself during the flipping procedure.
    TT, on the other hand, is vastly different and becomes pronounced for stiff rings: the polymer ring maintains its shape, but beads start moving
    along its shape around the vorticity axis. This can best be visualized by a coin rolling on its edge, or the wheels of a tank. 
   
    In previous work, we have highlighted the importance of fully-developed hydrodynamics when investigating ring polymers under shear \cite{liebetreu-acsml-2018}:
    Without rigidity, rings of sufficient length rarely show a pronounced TT motion, and short trefoil knots develop unique dynamics.
    Here, we present the emergence of an inflationary phase for ring polymers {\textit under simple, pure shear and fully-developed hydrodynamic interactions} 
    caused by a backflow from the horseshoe regions.
    In this phase, the ring undergoes full unfolding in all directions and transforms itself to an almost rigid, stretched, non-Brownian particle, in which 
    not only TT, but also TB motions are suppressed. The inflated regime is unique to ring polymers but independent of the presence
    of knots along their backbone but it requires a minimum polymer size to manifest itself. The aim of this work is to analyze the behavior of rings 
    with and without knotted topology, and to quantify and explain the interplay between hydrodynamic interactions and topology. 
    We suggest the swelling of the polymer in vorticity direction might potentially pave the way to distinguishing rings and chains of different sizes,
    and that shear could be used to reliably tighten knots on rings to unify the behavior of polymer melts. 
    The vorticity swelling of all rings and its massive impact on dynamics and shape 
    underlines the importance of fully-developed hydrodynamics when studying such polymers in solution.  


    \section*{Materials and Methods}
    We have employed Molecular Dynamics (MD) simulation coupled to a Multi-Particle Collision Dynamics (MPCD) solvent \cite{mpcd_original}, 
    together with Lees-Edwards boundary conditions \cite{lees-edwards-jpc-1972}, to simulate a variety of single ring polymers of various sizes
    and topologies under shear.
    In what follows, $N$ stands for the number of beads in the polymer; the topologies simulated were the unknot, $0_1$, as well as the prime knots
    $3_1$, $4_1$, $5_1$, $5_2$, $6_1$, $6_2$, $6_3$, $7_1$ and $8_1$, employing the Alexander-Briggs notation \cite{alexandbriggs} to characterize the knots.

    \bigskip\noindent
    \textbf{Multi-Particle Collision Dynamics.}
        The MPCD technique \cite{mpcd_original,mpcd_review} allows us 
        to simulate a particle-based, mesoscopic solvent with fully-developed hydrodynamic interactions. 
        MPCD features two alternating steps: streaming, where particles propagate, and collision, 
        where one separates the system into smaller cells of length $a$ and computes the center-of-mass velocity $\vb{v}_{\textrm CM}$ in each,
        performing a random rotation of all deviations of the cell-particle velocity vectors from the latter, and adding it back to produce new, post-collision
        velocities.
        Because our simulations involve shear, the system would continuously heat up by viscous heating. 
        To prevent this, we apply a cell-level Maxwellian thermostat \cite{mpcd_detailed}.
        We employ the usual choice of parameters \cite{liebetreu-acsml-2018, weiss-acsml-2017, starpolymers, mpcd_detailed} with a collision cell length $a$ 
        as a unit of length and rotation angle $\chi = 130^{\circ}$. All solvent particles are assigned a mass of $m$, serving as the unit of mass. 
        We set the average number of solvent particles per cell as $\langle N_{c} \rangle = 10$ and the time step $h = 0.1\,[(k_{\textrm B} T)^{-0.5} m^{0.5} a]$,
        with $k_{\textrm B}$ being Boltzmann's constant and $T$ the absolute temperature.
        With our choice of parameters, the solvent dynamic viscosity $\eta$ obtains the value \cite{mpcd_detailed} $\eta = \eta_{kin} + \eta_{col} \cong 8.70\,[(k_{\textrm B} T)^{0.5} m^{0.5} a^{-2}]$.

        \bigskip\noindent
        \textbf{Lees-Edwards boundary conditions.}
        The shear in our simulation is implemented by Lees-Edwards boundary conditions \cite{lees-edwards-jpc-1972} with a prescribed 
        shear rate $\dot\gamma$. With ${\vb{\hat{x}}}, {\vb{\hat{y}}}$ and ${\vb{\hat{z}}}$ being the flow, gradient and vorticity directions, respectively,
        these boundary conditions establish the velocity profile of planar Couette flow of the solvent in the absence of polymer as $\vb{v}_s({\textbf r}) = \dot\gamma y \vb{\hat{x}}$.

        \bigskip\noindent
        \textbf{Molecular Dynamics.}
        We employ the Kremer-Grest bead-spring model \cite{grest-pra-1986, kremer-jcp-1990} to investigate the properties of a fully-flexible polymer. 
        We employ $\epsilon = k_B T = 1.0$ and $\sigma = a = 1.0$ as units of energy and length from the Weaks-Chandler-Andersen (WCA) potential, respectively. 
        We set the parameters for the FENE potential to $k = 30\,\epsilon\sigma^{-2}$ and $R_0 = 1.5\,\sigma$.
        With these settings, the expected bond length $\langle l_b \rangle = 0.965\,\sigma$. 
        Velocity-Verlet \cite{velocityverlet_original, velocityverlet_explained} with a time step $\delta t = h / 100 = 0.001\,[(k_{\textrm B} T)^{-0.5} m^{0.5} a]$ was used to solve the MD-equations of motion.
        This algorithm is easily coupled to MPCD by setting the mass of each monomer $M = m \langle N_c \rangle$ and having the monomers participate in the 
        collision step \cite{mpcd_detailed, liebetreu-acsml-2018, starpolymers}.
        We have chosen to use the Alexander Polynomial \cite{alexander-tams-1928} to detect and the minimum-closure scheme to localize \cite{tubiana-ptps-2011} the knotted section on all simulated rings
        carrying knots. Simulating the various polymers in equilibrium allows us to determine as well their longest relaxation time $\tau_R$, from which the 
    Weissenberg number \cite{weissenbergorigin} $Wi = \dot\gamma\tau_R$ can be extracted for different contour lengths. A summary of the relaxation times
    is given in Table \ref{tab-relaxationtimes}.
    
    \begin{table}\centering
        \begin{tabular}{cccc}
            Knot type & HI & $N$ & $\tau_R\, [(k_{\textrm B}T)^{-0.5}m^{0.5}a]$ \\
            \midrule
            $0_1$ & $+$ & $100$ & $8543.84$ \\
            $0_1$ & $+$ & $150$ & $15758.60$ \\
            $0_1$ & $+$ & $200$ & $25902.20$ \\
            $3_1$ & $+$ & $100$ & $5104.43$ \\
            $3_1$ & $+$ & $200$ & $15874.20$ \\
            $4_1$ & $+$ & $200$ & $13720.90$ \\
            $5_1$ & $+$ & $200$ & $13219.00$ \\
            $5_2$ & $+$ & $200$ & $13826.10$ \\
            $6_1$ & $+$ & $200$ & $13404.70$ \\
            $6_2$ & $+$ & $200$ & $12012.00$ \\
            $6_3$ & $+$ & $200$ & $11673.50$ \\
            $7_1$ & $+$ & $200$ & $13669.80$ \\
            $8_1$ & $+$ & $200$ & $11394.30$ \\
            \bottomrule
        \end{tabular}
        \caption{Relaxation times $\tau_R = a_0 \tau_1 + (1 - a_0) \tau_2$ for a variety of different knot types, partly at different contour lengths. $a_0, \tau_1$ and $\tau_2$ are obtained from
        fitting $f(t) = a_0 \exp(- t / \tau_1) + (1-a_0) \exp(- t / \tau_2)$ to the autocorrelation function $\Phi(t) = {\langle \vb{R}_e(t) \cdot \vb{R}_e(0)\rangle}/{\langle \vb{R}^2_e \rangle}$.}
        \label{tab-relaxationtimes}
    \end{table}
        
    \bigskip\noindent
        \textbf{GPU and system size.}
        Our code has been written with CUDA/C++ \cite{nickolls-queue-2008-cuda-origin} to run in parallel on a GPU cluster \cite{westphal-cpc-2014-mpcd-gpu}. 
        This enabled more extensive simulations over a longer period of time to study flow profiles in detail.
        We also increased the simulation box volume to minimize residual effects by system size limitations. 
        Our box sizes were $V = (100 \times 60 \times 60) a^3$ for a number of monomers $N=100$ and $V = (150 \times 80 \times 80) a^3$ for $N=150$ and $N=200$. 
        We have tested the influence of the box size by running a simulation with $V = (200 \times 120 \times 120) a^3$ for $N=150$, which consistently reproduced the same behavior.        
        Boxes are elongated in the flow direction ${\vb{\hat{x}}}$ to accommodate the stretched polymer there, and quadratic in the gradient- and vorticity directions ${\vb{\hat{y}}}$ and ${\vb{\hat{z}}}$.



    \section*{Results}

    \textbf{Self-stabilization of long rings under shear and cessation of tumbling.}
    Contrary to open-ended topologies like chains and stars, hydrodynamic interactions on rings cause the polymer to swell not only in flow, 
    but also in the vorticity direction \cite{liebetreu-acsml-2018}. 
    To quantify the polymer's conformation under shear, we consider the gyration tensor \cite{rudnik-jpa-1986}:

    \begin{equation}
        G_{\alpha\beta} = \frac{1}{N} \sum_{i=1}^N \langle \vb{s}_{i,\alpha} \vb{s}_{i,\beta} \rangle,
    \end{equation}


    \begin{figure}[!ht]
        \centering
        \includegraphics{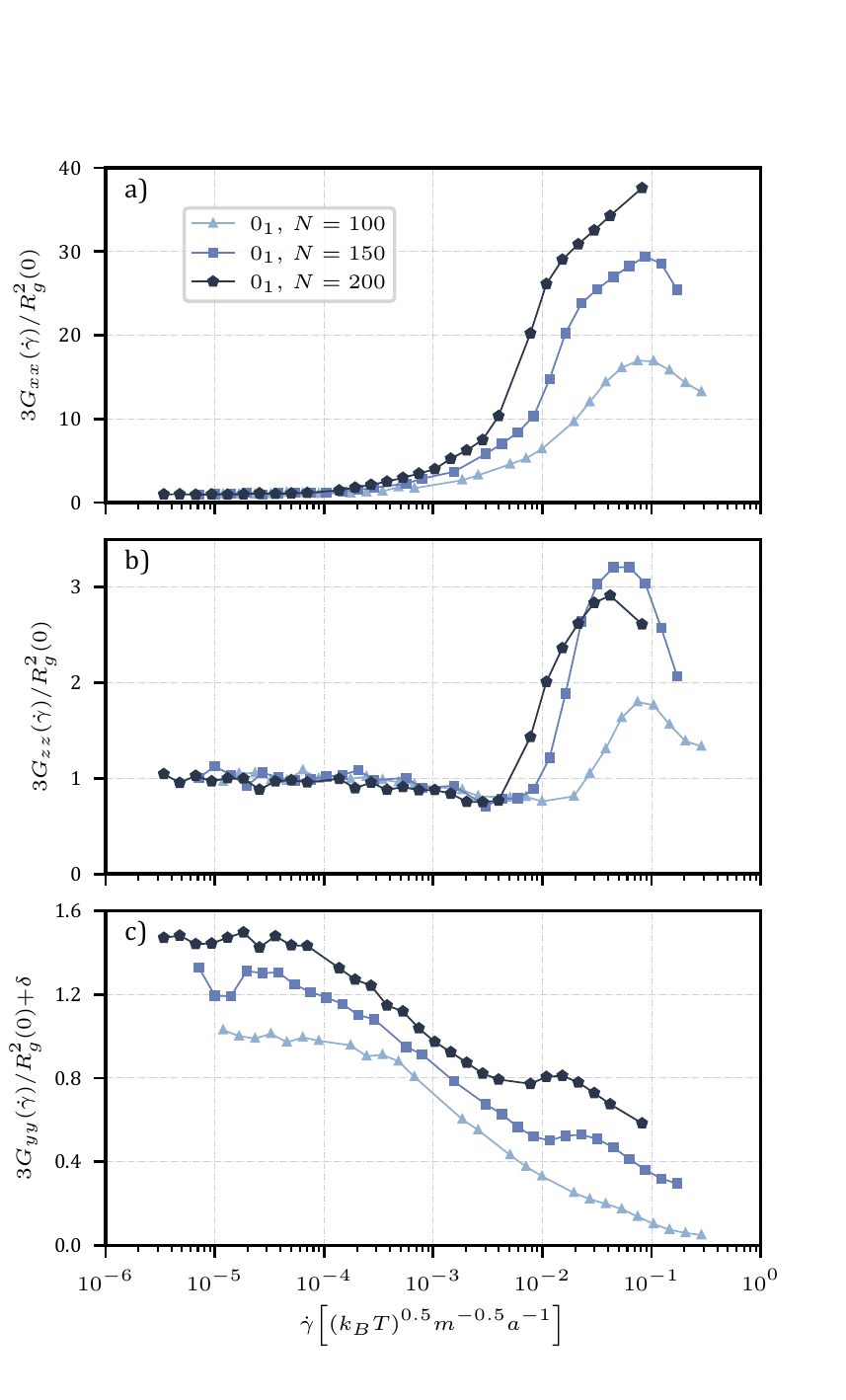}
        \caption{
            The gyration tensor diagonal elements $G_{\alpha\alpha}(\dot\gamma)$, normalized over their equilibrium value 
            $R_g^2(0)/3$, against the shear rate $\dot\gamma$. 
            $(\textrm{a})$: $G_{xx}$ (flow direction) increases with shear rate to some maximum and then drops as rings align into flow-vorticity plane and experience less strain. 
            $(\textrm{b})$: $G_{zz}$ (vorticity direction) decreases to a minimum and then shoots up at values of $\dot\gamma$ that
            anticipate the inflation anomaly observed in panel $(\textrm{c})$
            for the gradient direction.
            $(\textrm{c})$: $G_{yy}$ (gradient direction) initially decreases with shear rate. For $N=150$ and $N=200$, an inflation anomaly is observed, which is
            hardly visible for $N = 100$. 
            To highlight this behavior, shifts $\delta$ have been applied along the vertical axis, where $\delta_{N=100} = 0.0$, $\delta_{N=150} = 0.25$ and $\delta_{N=200} = 0.5$.
        }
        \label{fig1-gyrationdiagonals}
    \end{figure}

    where $\vb{s}_{i,\alpha}$ is the $\alpha$-coordinate of particle $i$'s position relative to the polymer's center-of-mass $\vb{\bar{r}}$, such that $\vb{s}_i = \vb{r}_i - \vb{\bar{r}}$
    and the symbol $\langle\cdots\rangle$ denotes the expectation value arising from performing an averaging over time. 
    The diagonal elements of this tensor express the polymer's extension in these directions and are shown in Fig.~\ref{fig1-gyrationdiagonals}, plotted against increasing shear rate $\dot\gamma$.
    With the help of the results in Table \ref{tab-relaxationtimes}, it can be established we have reached, for all polymers considered, Weissenberg numbers of order $10^3$,
    well into the strongly nonlinear regime, $Wi \gg 1$.
    
    In flow direction (Fig.~\ref{fig1-gyrationdiagonals}a), $G_{xx}$ increases steadily up to a certain shear rate of approximately $\dot\gamma \sim 0.1$ and then it starts decreasing.
    Along the vorticity direction (Fig.~\ref{fig1-gyrationdiagonals}b), $G_{zz}$ experiences a short decrease and then increases rapidly up to a maximum before decreasing again. 
    This effect is caused by backflow from the horseshoe regions of a stretched ring under shear and is exclusive to the ring topology \cite{liebetreu-acsml-2018}. 
    Already this non-monotonic behavior of the vorticity-direction diagonal component of the gyration tensor is quite unusual and unique to ring polymers
    but even more spectacular features show up in the gradient-direction diagonal element $G_{yy}$, shown in Fig.~\ref{fig1-gyrationdiagonals}c.
    Contrary to all other known polymer types, which display a monotonic decrease of $G_{yy}$ with the shear rate $\dot\gamma$, a swelling anomaly shows up for rings,
    resulting, for sufficiently long rings, in a non-monotonic behavior of $G_{yy}$ and featuring a local minimum, a rise to a local maximum and a further
    decrease of this quantity thereafter. The effect is almost invisible for a contour length of $N=100$, i.e., one needs sufficiently long rings to clearly
    identify it. 
         

    During the anomaly in the behavior of $G_{yy}$, the $N=150$ and $N=200$ rings undergo a shift in behavior, transitioning from the usual 
    motion that features strong thermal fluctuations on the monomer scale as well as tumbling of the whole macromolecule, towards a new phase in which the
    ring is stretched and unfolded, and the tumbling motion stops. At the end of this phase, which takes place for shear rates bracketed
    by the local minimum and the local maximum of $G_{yy}$, the ring behaves as a non-Brownian particle for which thermal fluctuations
    are strongly suppressed by the strong hydrodynamic and elastic forces between connected monomers. 
    Since this phase manifests itself as an unfolding of the polymer and a uniform expansion in all directions whilst the ring orientation
    in space remains fixed (see below) we term it \textit{inflation phase}, induced by shear and hydrodynamics. After the development
    of the inflation phase, the ring becomes fully inflated or stretched.

    Very recently, Young {\textit et al.}\ \cite{sing:pre:2019} applied Brownian Dynamics simulations of chains and rings in {\textit mixed} planar flows,
    approximating the effects of hydrodynamics via the Rotne-Prager-Yamakawa (RPY) mobility tensor. Though they did find a stretched phase for truly mixed flows
    that combine shear with extension, the inflation phase and the stretched configurations are absent in their findings for pure shear, 
    corresponding to the value $\alpha = 0$ of the velocity gradient tensor $\tilde{\Gamma}$ in the 
    standard terminology of mixed flows \cite{hur:pre:2002,woo:jcp:2003,sing:pre:2019}. 
    The absence of this phase in the aforementioned work could be a consequence of approximating hydrodynamics
    via the RPY-tensor but it is most likely an effect of the low degree of polymerization, $N = 120$, employed by Young {\textit et al.} Further differences with 
    Ref.\ \cite{sing:pre:2019} pertain to the solvent quality ($\theta$-solvent in Ref.\ \cite{sing:pre:2019} vs.\ good quality solvent here) and to the
    modelling of the bonding interactions between the monomers (harmonic vs.\ FENE-springs, respectively). These details should not play
    an important role at sufficiently high Weissenberg numbers, however.

\begin{figure*}[!ht]
        \centering
        \includegraphics{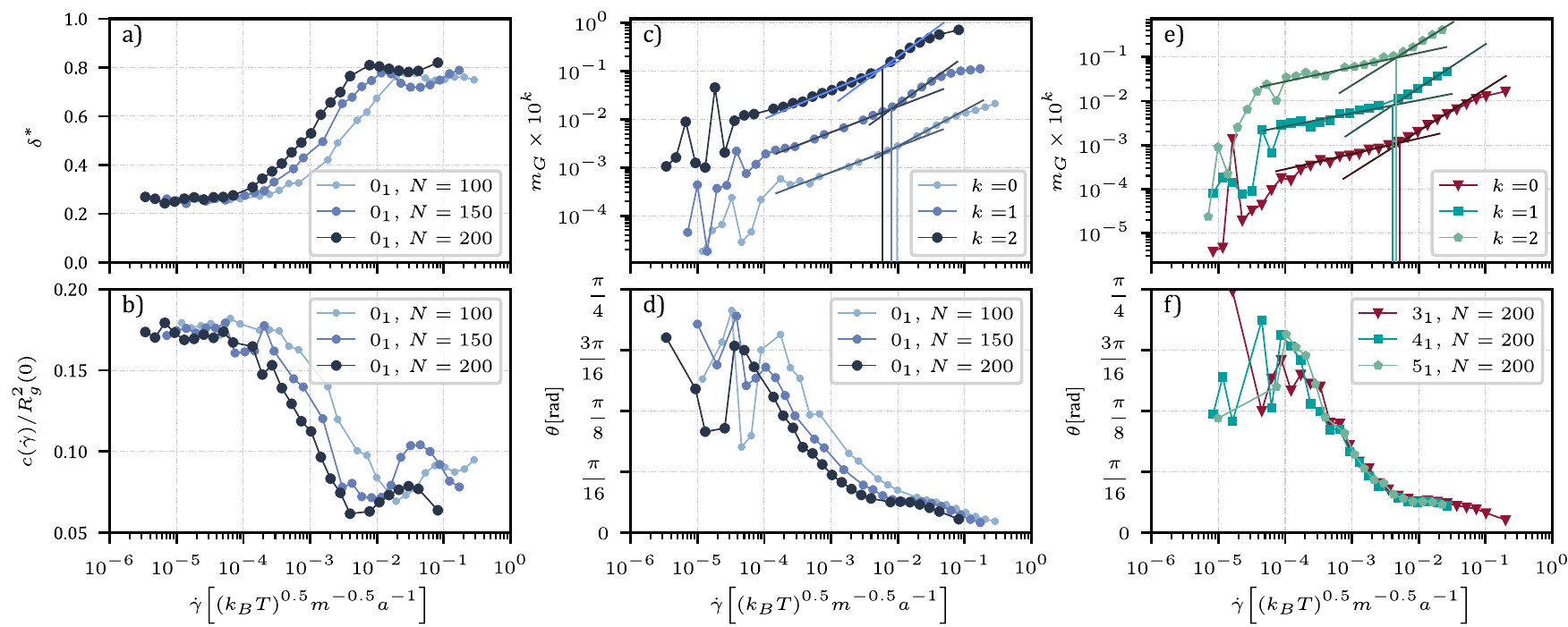}
        \caption{Shape and orientation of knotted and unknotted ring polymers under increasing shear rate $\dot\gamma$.
        $(\textrm{a})$: Anisotropy $\delta^{*}$. $(\textrm{b})$: Acylindricity $c(\dot\gamma)$ scaled over
        the squared gyration radius at $\dot\gamma = 0$. $(\textrm{c})$: Orientational resistance $m_G$
        for three different $0_1$-rings, as denoted in the label of panel $(\textrm{d})$. The vertical lines denote crossovers from the tumbling regime 
       at the lower values of $\dot\gamma$ to the inflation regime at higher ones. The straight lines on the left of the crossover have slope 0.6
       and the ones on the right have slope 0.85 for $N = 100$ and unity for $N = 150$ and $N = 200$. Curves have been multiplied by 
       constants, providing vertical shifts as indicated in the label. $(\textrm{d})$: Alignment angle $\theta$ between the 
       eigenvector corresponding to the largest eigenvalue of the gyration tensor and the flow axis for the three $0_1$-rings. 
       $(\textrm{e})$: Orientational resistance for three different knotted rings of topologies as denoted in the label of panel $(\textrm{f})$. 
       The straight lines on the left of the crossover points have slope 0.6 and the ones on the right have slope unity. The curves have been 
       multiplied by constants, whose values are given in the label, for visual clarity. $(\textrm{f})$: Same as panel $(\textrm{d})$ but for
       the three knotted rings whose characteristics are summarized in the label.}
        \label{fig2-shape}
    \end{figure*}

    The shape parameters \cite{rudnik-jpa-1986, aronovitz-jp-1986, ripoll-prl-2006, narros-mm-2013} are following the same anomalous behavior.
    Two of them, the asphericity and the acylindricity are shown as representatives in Figs.~\ref{fig2-shape}a and \ref{fig2-shape}b.  
    Indeed, the non-monotonic behavior of the elements of the gyration tensor is directly reflected in similar non-monotonic trends as the 
    aforementioned quantities. Very informative is the behavior of the orientational resistance $m_G = Wi\,\tan(2\theta)$ and of the
    alignment angle $\theta$ of the polymer with the flow axis, shown in Figs.~\ref{fig2-shape}c and \ref{fig2-shape}d, offering a direct
    quantitative measure of the onset and the characteristics of the inflation phase. It can be seen that $m_G$ shows a crossover from
    the usual, $m_G \sim \dot\gamma^{0.6}$-behavior to a much sharper,  $m_G \sim \dot\gamma$ power-law, which allows us 
    to define a characteristic crossover shear rate $\dot\gamma_{\times}$ marked by the vertical lines in Fig.~\ref{fig2-shape}c
    and summarized in Table~\ref{gammacross:tab}. The second regime immediately implies that the orientation angle $\theta$
    does not change with the shear rate $\dot\gamma$, see also the Discussion Section below. Moreover, as a direct comparison between
    Figs.~\ref{fig1-gyrationdiagonals}c, \ref{fig2-shape}c and \ref{fig2-shape}d shows, the regime between the local minimum and the
    local maximum of $G_{yy}(\dot\gamma)$ coincides with the regime of the $m_G \sim \dot\gamma$-scaling in which the orientation
    angle $\theta$ remains constant and the ring thus merely unfolds, growing in all directions -- precisely the inflation phase mentioned above.
    Figs.~\ref{fig2-shape}e and \ref{fig2-shape}f show that the same scenario also holds for some selected knotted topologies, to which we will return in the
    next subsection.
            

    \begin{figure*}[!ht]
        \centering
        \includegraphics{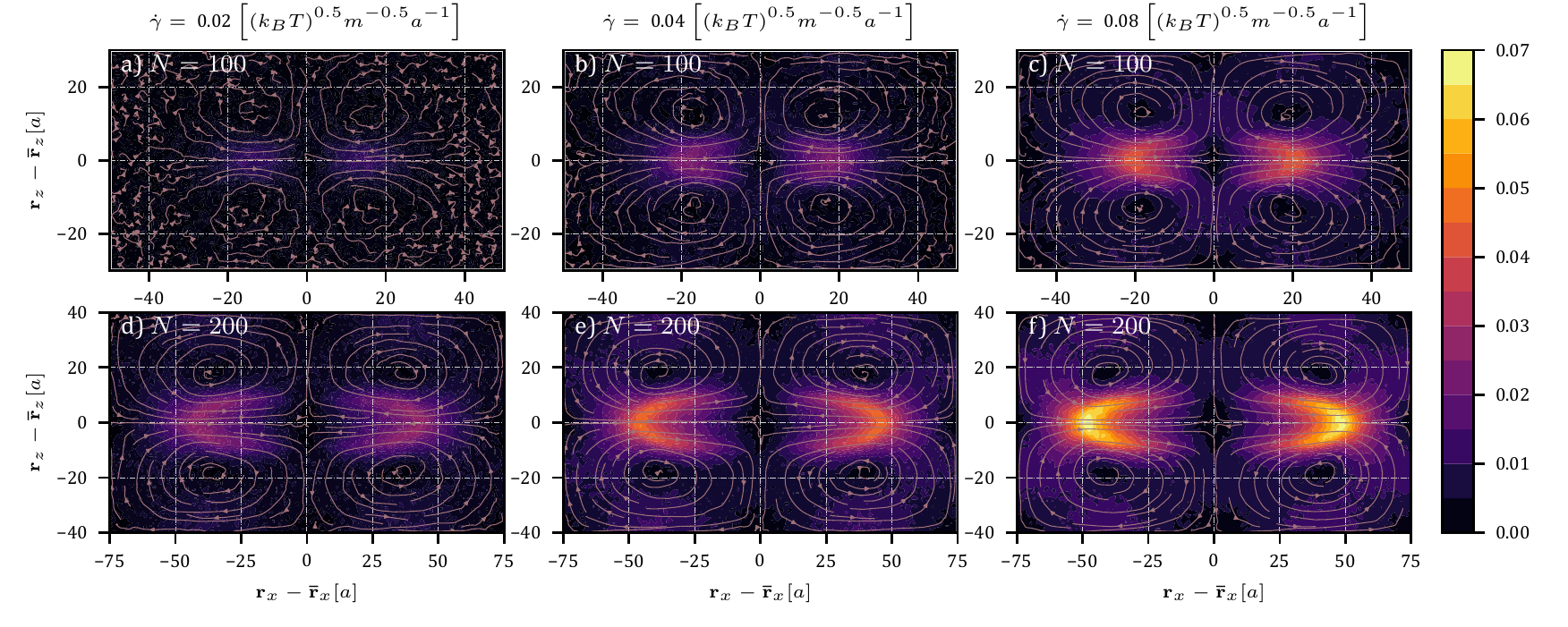}
        \caption{Flow fields in flow-vorticity plane centered around center of mass $\vb{\bar{r}}$ of an $0_1$ ring. 
        The colour map encodes velocity magnitudes in units of $[\sqrt{k_B T / m}]$. 
        Left to right: Increasing shear rates $\dot\gamma = {0.02, 0.04, 0.08} ~[(k_B T)^{0.5} m^{-0.5} a^{-1}]$.
        Top to bottom: Increasing contour length $N = {100, 200}$. 
        Direct comparison between equal shear rates shows backflow is significantly more pronounced at $N=200$.
        }
        \label{fig3-flowfields}
    \end{figure*}

\begin{table}
    \begin{center}
    
    \begin{tabular}{ccc}
        Ring topology & $N$ & $\dot{\gamma}_{\times}\,\, [(k_B T)^{0.5} m^{-0.5}a^{-1}] $ \\
        \midrule
$0_1$ & $100$ & $9.8 \times 10^{-3}$ \\
        $0_1$ & $150$ & $8.0 \times 10^{-3}$ \\
        $0_1$ & $200$ & $5.9 \times 10^{-3}$ \\
        $3_1$ & $100$ & $1.3 \times 10^{-2}$ \\
        $3_1$ & $200$ & $5.1 \times 10^{-3}$ \\
        $4_1$ & $200$ & $4.0 \times 10^{-3}$ \\
        $5_1$ & $200$ & $4.7 \times 10^{-3}$ \\
        $5_2$ & $200$ & $6.2 \times 10^{-3}$ \\
        $6_1$ & $200$ & $6.6 \times 10^{-3}$ \\
        $6_2$ & $200$ & $7.2 \times 10^{-3}$ \\
        $6_3$ & $200$ & $6.4 \times 10^{-3}$ \\
        $7_1$ & $200$ & $7.7 \times 10^{-3}$ \\ 
        $8_1$ & $200$ & $7.9 \times 10^{-3}$ \\       
        \bottomrule
    \end{tabular}\\
\end{center}
        \caption{Crossover Shear Rates. The crossover values $\dot\gamma_{\times}$ for the transition from the tumbling to the inflation phase of ring polymers with
        topology denoted at the left column and degree of polymerization $N$ in the middle one. Note that for the cases of 
        $N = 100$ ($0_1$ and $3_1$-rings), a fully developed inflation phase does not emerge and $\dot\gamma_{\times}$ denotes there the crossover value between the
        two power-law regimes in Fig.~\ref{fig2-shape}(b).}
        \label{gammacross:tab}
\end{table}

\begin{figure*}
    \centering
    \includegraphics[width=\linewidth]{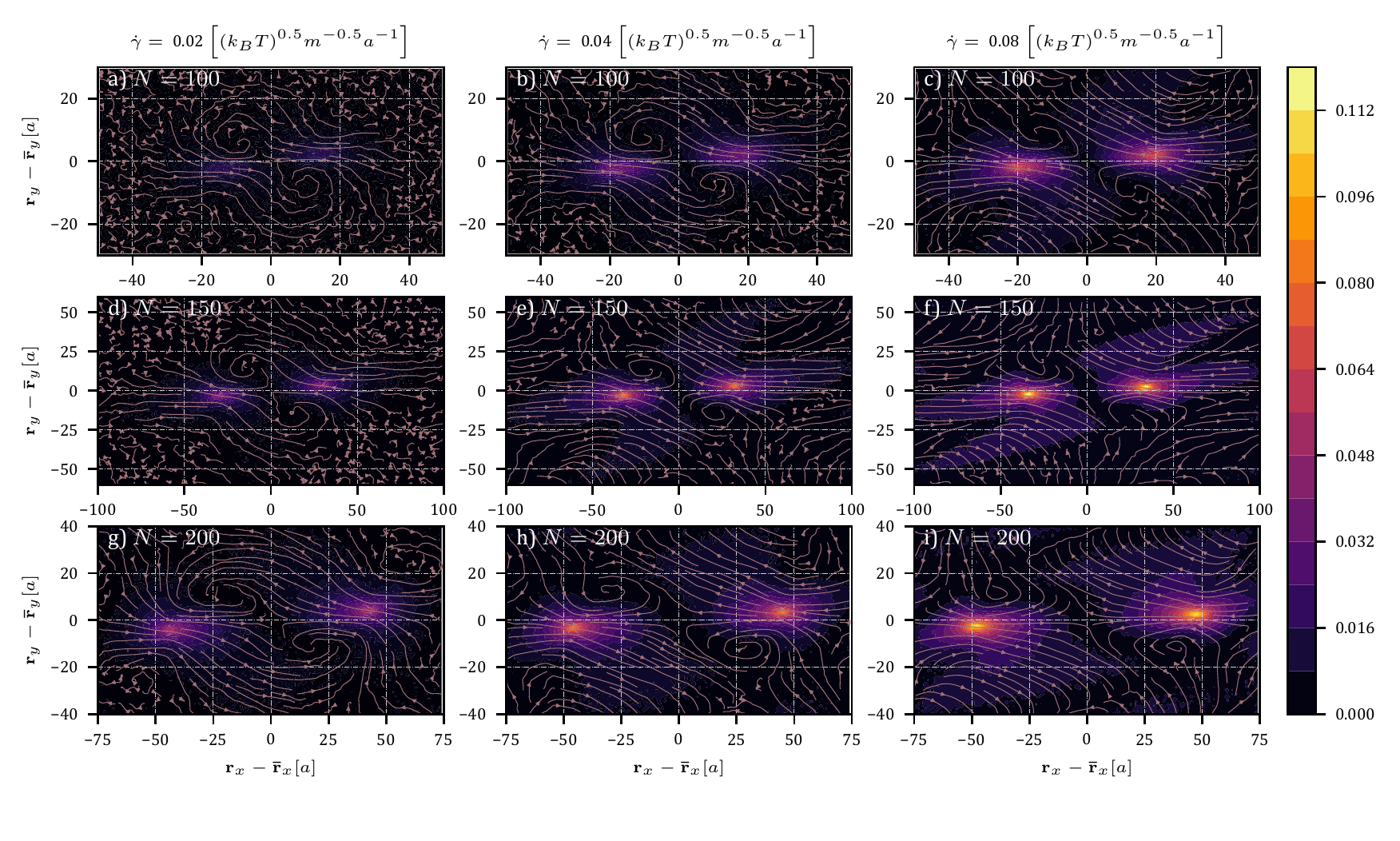}
    \caption{Flow fields in flow-gradient plane centered around center of mass $\vb{\bar{r}}$ of an $0_1$ ring. 
    The colour map encodes modified velocity magnitudes $\|\vb{v}'\|$ in units of $[\sqrt{k_B T / m}]$, $\vb{v}' = \vb{v} - \dot\gamma y \vb{\hat{x}}$.
    Left to right: Increasing shear rates $\dot\gamma = [0.02, 0.04, 0.08] ~[(k_B T)^{0.5} m^{-0.5} a^{-1}]$.
    Top to bottom: Increasing contour length $N = [100, 150, 200]$.}
    \label{fig-si2-flowfields-fg-pure}
\end{figure*}

\begin{figure*}
    \centering
    \includegraphics[width=\linewidth]{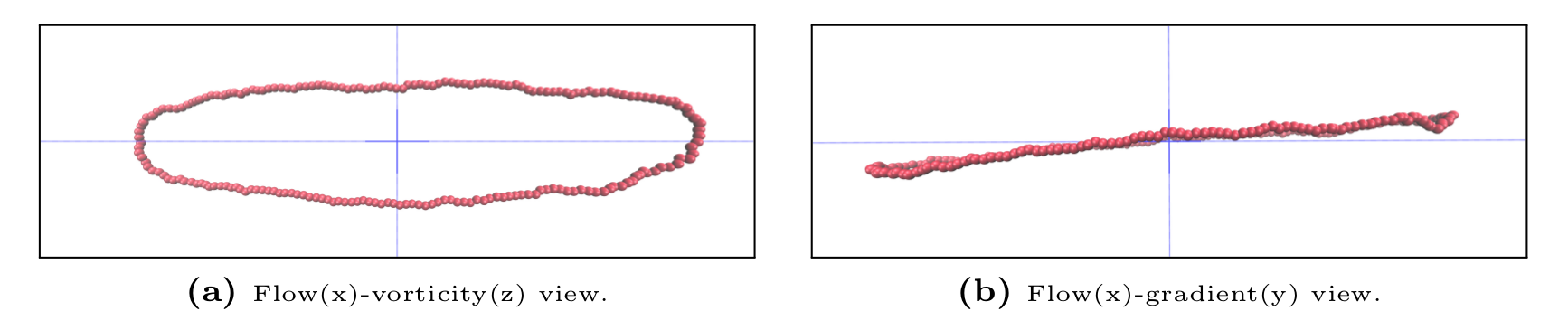}
    \caption{Typical and stable swollen configuration of the $N=200$ $0_1$-ring, at $Wi=543.95$ or $\dot\gamma = 2.1 \times 10^{-2} \left[(k_BT)^{0.5} m^{-0.5} a^{-1}\right]$.
    We have marked the flow (x) and vorticity (z) or gradient (y) axis in blue. The intersection is the polymer's center of mass.
    The solid part on each axis extends a length of $5 [a]$ in either direction.}
    \label{fig-si6-01-n200-stable}
\end{figure*}

    To further describe, analyze
    and understand shear-induced ring polymer inflation, 
    we take a closer look at the $N=200$ ring as an example:
    \begin{itemize}
        \item \textbf{In the tumbling regime:} At $\dot{\gamma} \cong 2.5 \times 10^{-3}$, $G_{yy}$ first reaches the value of the anomaly maximum. 
        We observe the formation of a tilting axis along which the polymer stretches. Tumbling frequently occurs and takes up substantial space in gradient direction.
        At $\dot\gamma \cong 3 \times 10^{-3}$, tumbling becomes less gradient-intense and occurs less frequently, 
    but the tilting becomes more pronounced as fluctuations around the axis decrease.
        \item \textbf{During the inflation phase:} 
        Before the anomaly maximum of $G_{yy}$, at $\dot\gamma \cong 8 \times 10^{-3}$, there are rare tumbling events as the ring starts to stabilize and open. Less fluctuations are visible. 
        \item \textbf{In the fully inflated regime:} At $\dot\gamma \cong 1.5 \times 10^{-2}$, the anomaly reaches its peak. The polymer is aligned with the tilting axis and quite stable. Tumbling is extremely rare.
        Finally, at $\dot\gamma \cong 2 \times 10^{-2}$, the polymer does not tumble but looks almost rigid, while the tilting axis gets progressively closer to the flow axis.         
    \end{itemize}
    
    During the inflation
    phase and all the way into the fully inflated configuration, a particular kind of tank-treading (TT) motion shows up in lieu of the suppressed tumbling. Contrary to the usual 
    shear-induced TT in which the rotation axis lies parallel to the vorticity direction, here we have fluctuation-induced rigid rotations of the inflated molecule around an 
    axis lying in the flow-gradient direction perpendicular to its tilt, i.e., almost parallel to the gradient direction since $\theta$ is small. 
    
    To visualize the contribution of the solvent, in conjunction with the closed polymer topology,
    to this behavior, we first compare the flow profiles established in the flow-vorticity plane around the polymer's center-of-mass for contour lengths $N=100$ and $N=200$
    as shown in Fig.~\ref{fig3-flowfields}. Qualitatively, the established flow fields are similar, but at the same shear rates, the $N=200$ ring experiences a stronger backflow,
    and the shape and position of the horseshoe is more visible in the flow profile. At the maximum of the anomaly on $G_{yy}$ at around $\dot\gamma = 0.02$ (Fig.~\ref{fig1-gyrationdiagonals}c),
    Fig.~\ref{fig3-flowfields}d shows a clearly established backflow profile, while the same cannot be said about the $N=100$ ring at the same shear rate (Fig.~\ref{fig3-flowfields}a).

\begin{figure*}[!ht]
        \centering
        \includegraphics{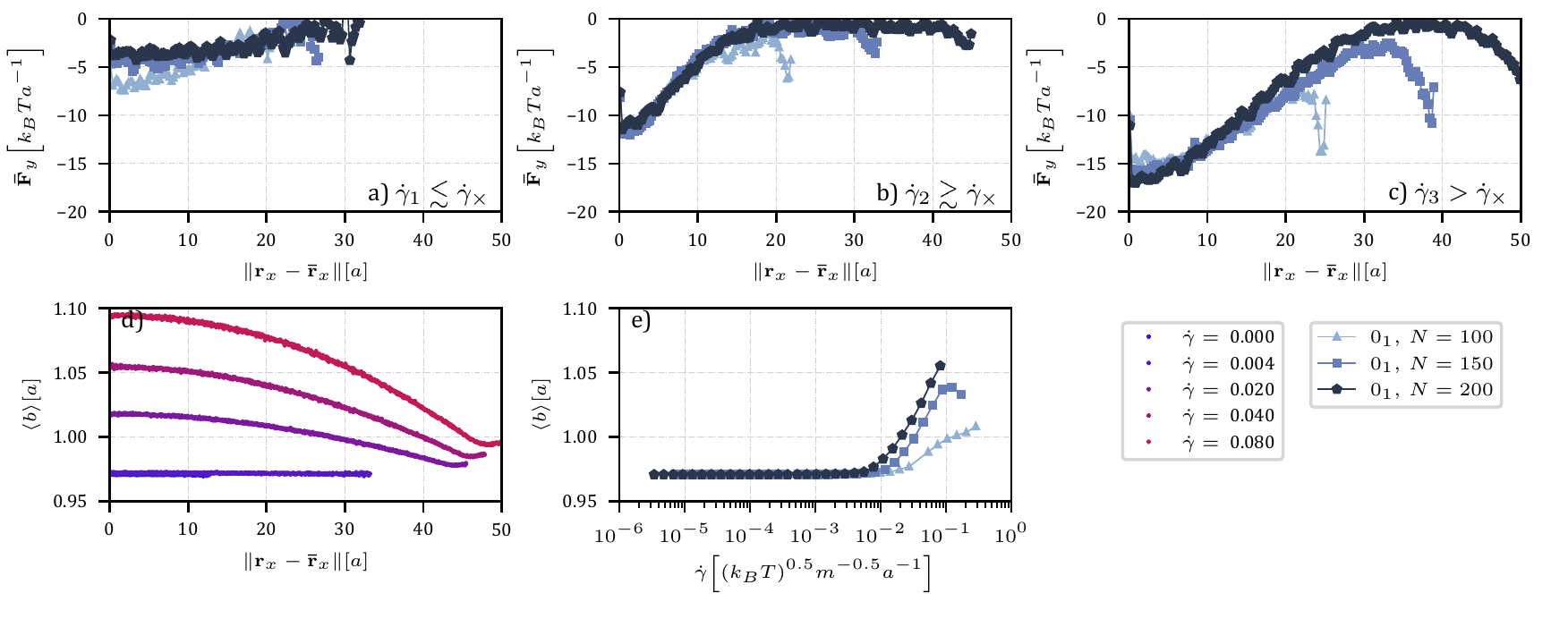}
        \caption{Inter-monomer forces $(\textrm{a-c})$ and bond extension along the ring $(\textrm{d-e})$. 
        $(\textrm{a-c})$: Gradient-direction component of force on beads, averaged in intervals along flow axis and measured relative to center-of-mass. 
        A negative sign signifies pull towards center-of-mass. Shear rate increasing from $(\textrm{a})$ to $(\textrm{c})$. 
        $(\textrm{a})$: At $\dot\gamma_1 = \left[0.020, 0.006, 0.003\right]$ for $N=[100, 150, 200]$, $G_{yy}$ first takes on the value of its local maximum. 
        This has to be estimated for $N=100$, where monotony is maintained.
        $(\textrm{b})$: At $\dot\gamma_2 = \left[0.040, 0.020, 0.018\right]$ for $N=[100, 150, 200]$, $G_{yy}$ reaches its local maximum.
        $(\textrm{c})$: At $\dot\gamma_3 = \left[0.286, 0.171, 0.082\right]$ for $N=[100, 150, 200]$, $G_{yy}$ has gotten substantially smaller than it was at the local maximum.
        $(\textrm{d})$: Bond extension of the $N=200$ ring along flow axis. $0.97~[a]$ is the average for equilibrium with the parameters given in the simulation. 
        $(\textrm{e})$: Average bond extension increasing with shear rate. 
        }
        \label{fig4-forces}
    \end{figure*}

    The transition to the inflated phase is closely related to the pattern of inter-monomer and hydrodynamic forces acting on the ring.
    Due to the development of a strongly stretched configuration, the former are exclusively elastic forces from the tethering (FENE) 
    potentials between successive beads; contacts between monomers are extremely rare and excluded volume interactions are inactive there. 
    In Figs.~\ref{fig4-forces}(a-c) we show the gradient-direction component of the inter-monomer
    force, $\vb{\bar{F}}_y$, as a function of the distance from the center of mass for three different ring sizes and shear rates slightly below,
    slightly above and well above the crossover value $\dot\gamma_{\times}$ for each length. 
    The observed force is always directed towards the center-of-mass and is strongest near the latter,
    where also the tension along the chain is strongest, as can be seen in Fig.~\ref{fig4-forces}d. The gradient-component forces feature, for the two longest
    rings that undergo an inflation phase, a striking development as $\dot\gamma_{\times}$ is crossed:
    a prolonged region along the ring where $\vb{\bar{F}}_y \cong 0$ develops, which arises from the emergence of portions of the
    ring that remain essentially horizontal, so that no $y$-component of the force results, see Fig.~\ref{fig-si6-01-n200-stable}b. At the same time, the
    fluid streams also smoothly and almost horizontally along these portions, see Figs.~\ref{fig-si2-flowfields-fg-pure}e and \ref{fig-si2-flowfields-fg-pure}h, so that the ring is 
    at stable equilibrium there. At the tips of the ring, as well as close to their centers, the inter-monomer forces tend to push the beads
    towards the neutral plane and they are counteracted by hydrodynamic forces of the streaming solvent, which is deflected on the 
    polymer, see Fig.~\ref{fig-si2-flowfields-fg-pure}. As we move to higher shear rates, this region gets smaller, and there are stronger forces acting on beads far away from the center-of-mass. 
    
    Referring to the typical shear rates $\dot\gamma_1$,  $\dot\gamma_2$, and $\dot\gamma_3$
    in Figs.~\ref{fig4-forces}(a-c), we propose the following 
    response patterns to shear for rings of sufficiently large contour length. 
    As shear rate increases from equilibrium, ring polymers start swelling first in flow direction and align with an axis in the flow-gradient plane. 
    They fluctuate around this axis, and occasionally experience tumbling because of sufficiently strong fluctuations of the horseshoe regions. 
    Approaching $\dot\gamma_1$, tumbling becomes enhanced and the knot starts to experience backflow from the horseshoe regions.
    Near $\dot\gamma_2$, tumbling stops completely. The backflow causes so much tension along the ring that fluctuations are suppressed. 
    For tumbling to occur, an entire section of the ring would have to angle towards the center-of-mass plane. 
    At the same time, during the transition between $\dot\gamma_1$ and $\dot\gamma_2$, the alignment angle is almost constant, so the extension of the 
    ring in gradient direction, and by extension the strength of shear and backflow, scales proportional to shear rate. 
    Eventually, transitioning from $\dot\gamma_2$ to $\dot\gamma_3$, the sheared solvent flows powerfully enough to push the ring 
    into the flow-vorticity plane, and the hydrostatic bubble becomes less and less pronounced. 
    In contrast, rings with low contour length never experience a point where fluctuations are suppressed enough to prevent tumbling,
    and a local maximum for $G_{yy}$ does not show up. 

 \begin{figure*}[!ht]
        \centering
        \includegraphics{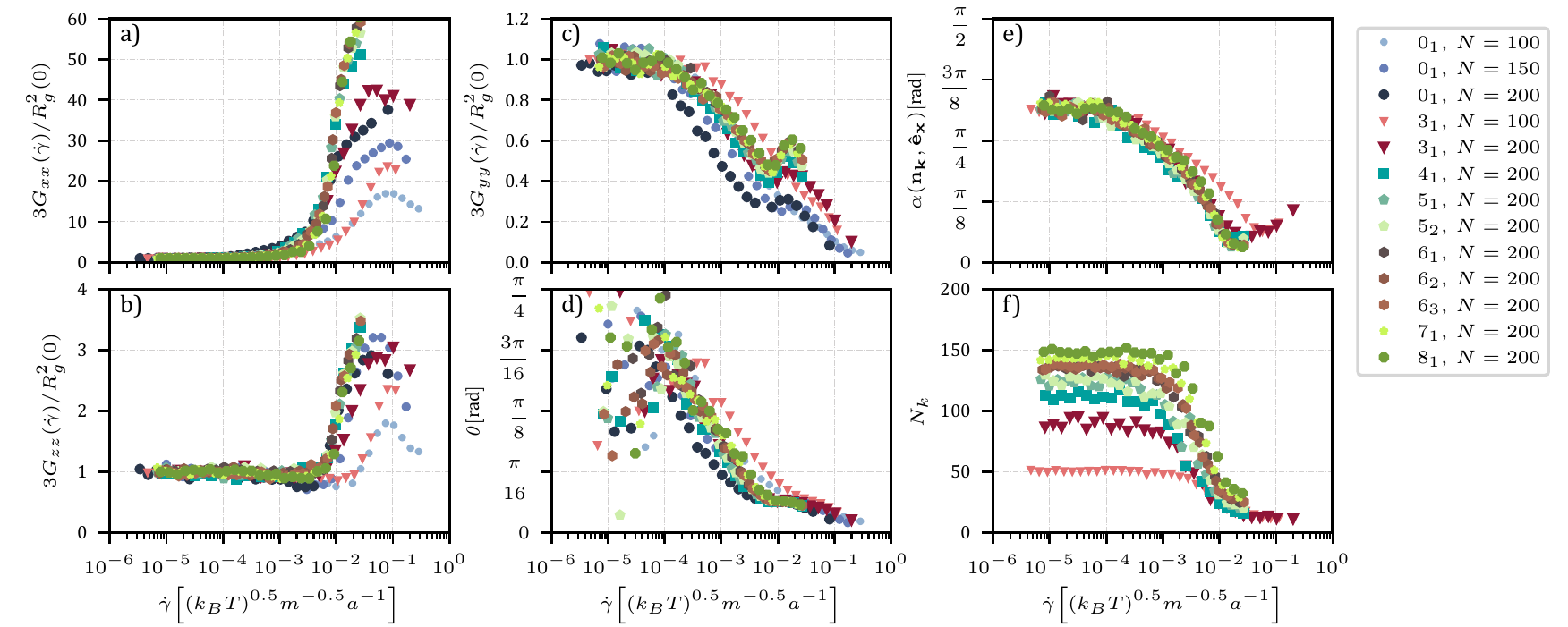}
        \caption{Shape and orientation of knotted ring topologies under increasing shear rate $\dot\gamma$. 
        $(\textrm{a})$: $G_{xx}$ (flow), $(\textrm{b})$: $G_{zz}$ (vorticity), $(\textrm{c})$: $G_{yy}$ (gradient), $(\textrm{d})$: alignment angle $\theta$,
        $(\textrm{e})$: angle $\alpha$ between vector $\vb{n_k}$ from the ring's to the knotted section's center of mass and axis vector $\vb{\hat{x}}$,
        and $(\textrm{f})$: number of beads on the knotted section, $N_k$. Under strong shear, all knot topologies behave almost exactly like a ring with the same contour length $N$.
        Different colors correspond to different knot types, as indicated in the legend.}
        \label{fig5-knots}
    \end{figure*}
    

\begin{figure*}[!ht]
        \centering
        \includegraphics[width=17.8cm]{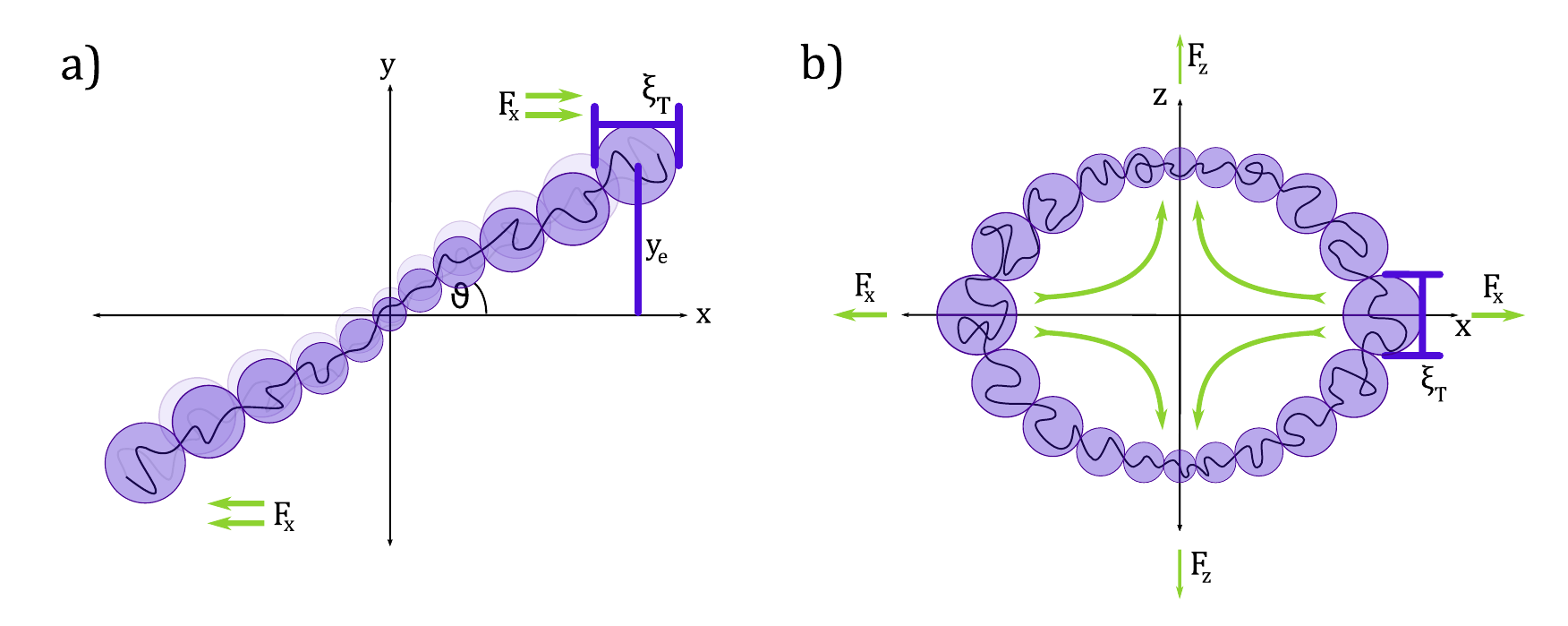}
        \caption{Sketch of a ring polymer under steady shear.
        $(\textrm{a})$: Viewed projected on the flow-gradient ($x$-$y$) plane, the ring is rendered as a succession of blobs, the largest of which 
        is located at the tips and has size $\xi_{\textrm T}$. The elevation of the tip over the neutral (flow-vorticity) plane is $y_{\textrm e}$ and the tilt angle 
        is $\theta$. 
        $(\textrm{b})$: Viewed projected onto the flow-vorticity ($x$-$z$) plane, the ring has an oval shape and it is stretched outwards in all
        directions due to backflow of solvent (denoted by the curved green arrows), reflected at the two horseshoe-shaped regions at the tips.
        }
        \label{fig6-tensionblobs}
\end{figure*}

\bigskip\noindent
    \textbf{Collapse of all knots and localization in space.} We have also investigated and compared behavior of knotted rings and report very similar behavior under strong shear. 
    The diagonal entries of the gyration tensor 
    (Fig.~\ref{fig5-knots}a-c) all show the same qualitative behavior, and across all knot topologies, and the anomaly is present for $G_{yy}$ (Fig.~\ref{fig5-knots}c). 
    The stabilization of the alignment angle is common across all topological varieties as well (Fig.~\ref{fig5-knots}d), and we conclude that shape-wise, it is justified to talk 
    about a generic behavior of ring-shaped topologies rather than just of the $0_1$ ring.

    All knotted sections are pulled tight (Fig.~\ref{fig5-knots}f) and stay small for the remainder of the simulation, essentially filling the role of a particularly bulky bead.
    This bead-like knotted section has a preferred position on the ring, although this constraint might be lifted as the ring aligns more and more with the flow axis. 
    Below this threshold, however, the knot tends to be located on the horseshoe-region of the ring (Fig.~\ref{fig5-knots}e) and, as tumbling is suppressed, 
    only has tank-treading as a possible way to move swiftly. In this sense, the knotted section acts as a stabilization anchor for the ring, which already experiences
    a strongly suppressed, fluctuating form of tank-treading rather than a pronounced one \cite{liebetreu-acsml-2018}. The knotted section adds to this effect.
    When the knotted section does move, and when it manages to leave the horseshoe region, it very quickly tank-treads to the opposite horseshoe region to
    stabilize there again. 
    
    \section*{Discussion}
    
    \textbf{Interpretation of results.} To understand the physical mechanism behind the transition from the tumbling to the inflation phase and to make quantitative predictions
    about its occurrence, we need to properly understand the interplay between polymer orientation in the shear field and the hydrodynamic 
    forces imposed on it by the solvent. In Fig.~\ref{fig6-tensionblobs}(a) the average tilt angle $\theta$ of the ring is shown, subtended by the
    vector corresponding to the largest eigenvalue of the gyration tensor and the flow axis, alongside with the elevation $y_{\textrm e}$ of the tip at the 
    horseshoe region of the ring. In Fig.~\ref{fig6-tensionblobs}(b) a top view of the ring is shown, featuring an open ellipsoidal shape characteristic
    of the swelling in the vorticity direction, caused by the backflow of the solvent reflected in the horseshoe regions \cite{liebetreu-acsml-2018}.
    Due to the form of the flow pattern, shown in Fig.~\ref{fig3-flowfields} and sketched in Fig.~\ref{fig6-tensionblobs}(b) by the curved green arrows,
    the whole ring experiences an outwards pointing inflation force (akin to the pressure in the interior of an elastic bubble) and thus it can be 
    thought of as a succession of tension blobs. The largest of these, having size $\xi_{\textrm T}$, is located at the tips of the ring. A necessary 
    condition for tumbling is 
    \begin{equation}
    \xi_{\textrm T} \geq y_{\textrm e},
    \label{xit:eq}
    \end{equation}
   so that parts of the upper (lower) tip of the ring can be found below (above) the neutral plane and carried to the opposite direction by the 
   streaming fluid. Accordingly, we need to estimate the dependence of  $y_{\textrm e}$ and $\xi_{\textrm T}$ on the shear rate $\dot\gamma$.
       
   From geometrical considerations, we have 
   \begin{equation}
   y_{\textrm e} \cong \sqrt{G_{xx}}\tan\theta \cong \sqrt{G_{xx}}\,\theta,
   \label{ye:eq}
   \end{equation}
   for small values of $\theta$, typical of the transition to the inflation phase.
   The flow-direction component of the gyration tensor,
   $G_{xx}$ scales in the tumbling regime as $G_{xx}(\dot\gamma)/G_{xx}(0) \sim {Wi}^{0.6}$ \cite{liebetreu-acsml-2018,chen-polymer-2015}. Keeping in mind that
   $G_{xx}(0) \propto R_g^2(0) \sim N^{2\nu}$ with the Flory exponent $\nu \cong 0.588$, the above considerations yield
   \begin{equation}
   \sqrt{G_{xx}} \sim \dot\gamma^{0.3}\tau_R^{0.3}\,N^{\nu}.
    \label{gxx:eq}
    \end{equation}
    The dependence of the tilt angle $\theta$ on the shear rate can be extracted from the scaling of the orientational resistance $m_G = Wi\tan(2\theta) \cong 2\theta\, Wi$
    on the Weissenberg number: $m_G \sim Wi^{0.6}$, implying
    \begin{equation}
    \theta \sim \dot\gamma^{-0.4}\,{\tau_R}^{-0.4}.
    \label{theta:eq}
    \end{equation}
    Combining Eqs.\ (\ref{ye:eq}), (\ref{gxx:eq}) and (\ref{theta:eq}) above, together with the power-law $\tau_R \sim N^{3\nu}$
    for the dependence of the longest relaxation time of the polymer on $N$, we obtain
    \begin{equation}
    y_{\textrm e} \sim \dot\gamma^{-0.1}\,N^{0.4},
    \label{yescale:eq}
    \end{equation}
    where the approximation $0.7\nu \cong 0.4$ has been employed. 
    
    The magnitude of the inflating force $F$ that stretches the ring
    and gives rise to the tension blobs is set by the streaming velocity of the solvent at the point of highest elevation $y_{\textrm e}$ and
    it scales as $F \sim \dot\gamma\,y_{\textrm e}$. The ensuing tension blob size scales as $\xi_{\textrm T} \sim k_{\textrm B}T/F \sim \dot\gamma^{-1}y_{\textrm e}^{-1}$, resulting,
    together with Eq.\ (\ref{yescale:eq}) above, into
    \begin{equation}
    \xi_{\textrm T} \sim \dot\gamma^{-0.9}\,N^{-0.4}.
    \label{xiscale:eq}
    \end{equation}
    Evidently, $\xi_{T}$ decreases with the shear rate $\dot\gamma$ much faster than $y_{\textrm e}$, so that beyond a crossover shear rate 
    $\dot\gamma_{\times}$, the inequality (\ref{xit:eq}) cannot be fulfilled and tumbling stops, the ring transitioning into the inflation phase.
    Putting together Eqs.\ (\ref{xit:eq}), (\ref{yescale:eq}), and (\ref{xiscale:eq}), we find the dependence of $\dot\gamma_{\times}$ with $N$ to follow a simple power-law: 
    \begin{equation}
    \dot\gamma_{\times} \sim \frac{k_{\textrm B}T}{\eta\sigma^3}N^{-1},
    \label{gammadotcrossover:eq}
    \end{equation}
   where we have reintroduced in the prefactor the quantities entering in the expressions of $y_{\textrm e}$ and $F$, to obtain
   the full dependence on the temperature, solvent viscosity and monomer size as well. 
   
   There is a striking comparison than
   can be made here with the case of suspensions of colloidal particles of size $d$ under shear. There, 
   the Peclet number $Pe$, which scales as
   \begin{equation}
   Pe \sim \frac{\eta\dot\gamma d^3}{k_{\textrm B}T},
   \label{peclet:eq}
   \end{equation}
   expresses the typical ratio of timescales for Brownian and shear-induced motions \cite{denn:arcbe:2014}. The Brownian character of the motion
   dominates for $Pe \lesssim 1$ whereas shear takes over for $Pe > 1$ and the particles become fully non-Brownian for 
   $Pe \gg 1$. Accordingly, and defining a crossover value $Pe_{\times} = 1$ between the two regimes, this translates, for colloidal
   particles, into a crossover shear rate
   \begin{equation}
    \dot\gamma_{\times} \sim \frac{k_{\textrm B}T}{\eta d^3}.
    \label{gammadotcrossovercolloids:eq}
    \end{equation}
    Comparing Eqs.\ (\ref{gammadotcrossover:eq}) and (\ref{gammadotcrossovercolloids:eq}), we see that the former is fully consistent with the interpretation and the physical picture of the inflated,
    stressed ring as a succession of $N$ particles of size $\sigma$ held together by essentially rigid connections. The fact that it has been
    derived independently, using scaling arguments from polymer theory, and at the same time it turns out to deliver a physically consistent picture of an inflated
    polymer as a rigid, non-Brownian colloid, offers additional corroboration of its validity.
   
   Using the typical orders of 
   magnitude, $\eta \cong 10$ and $N \cong 10^2$ employed in this work, we obtain $\dot\gamma_{\times} \cong 10^{-3}$,
   in satisfactory agreement with the results in Table \ref{gammacross:tab}. Though we have only simulated rings with three different $N$-values, 
   the results shown in Fig.~\ref{fig2-shape}(b) and summarized in Table \ref{gammacross:tab} also support the above power-law prediction. 
   Moreover, translating the crossover shear rate into a crossover Weissenberg number $Wi_{\times} = \dot\gamma_{\times}\tau_R$ and using the
   scaling $\tau_R \sim N^{3\nu}$, we obtain $Wi_{\times} \sim N^{0.76}$. Accordingly, employing ring polymers of high $N$ 
   guarantees that the condition of lying in the strongly nonlinear regime, $Wi_{\times} \gg 1$, is satisfied. 
   At the same time, increasing $N$ lowers the value $\dot\gamma_{\times}$ of 
   the shear rate at which the transition to the inflation phase will take place, therefore making the phenomenon more easily 
   observable experimentally.
    
    Once the rings
    enter their inflation phase, the orientational resistance scales as $m_G \sim Wi$, directly implying $\theta \sim Wi^0$: the orientation angle remains
    constant during this phase and the ring simply fully unfolds until the tension blob reaches the monomer size. Thereafter, it behaves as a rigid,
    non-Brownian object for which thermal fluctuations play no role. The highest elevation scales as $y_{\textrm e} \sim N$ [cf.\ Eq.\ (\ref{yescale:eq})] and the 
    hydrodynamic force on the monomers scales, accordingly, as $F \sim \dot\gamma\,N$, cf.\ $F \sim \dot\gamma^{0.9}\,N^{0.4}$ in the tumbling regime.
   
    \bigskip\noindent
    \textbf{Filter for chains and rings.} The swelling of rings in vorticity direction is exclusive to their closed structure and their transition to the stretched,
    inflated configuration is size-specific. Accordingly, we surmise that this property could be employed in a simple
    microfluidic device to enhance separation of large and small rings in a mixture under Poiseuille flow. Indeed, the
    latter creates a parabolic velocity profile for the Newtonian solvent in which the mixture is suspended, causing thereby
    a position-varying shear rate $\dot\gamma(y)$ across the gradient direction $y$. This local shear rate has a maximum
    close to the walls and vanishes in the middle of the channel. Large rings exposed to high shear close to the walls
    will inflate and stretch, orient themselves almost parallel to the wall and thus be subject to both steric constrains
    with one another and strong lift forces from the wall. It is thus natural to expect that they will migrate towards the
    center of the channel, and displacing from there smaller rings that will be pushed towards the walls. In this way, 
    a flow-induced size separation via focusing of the large rings in the middle can be achieved. The same is expected
    to be true for linear-ring mixtures.

    \section*{Conclusion}
    
    We have shown that ring polymers in the presence of fully-developed hydrodynamic interactions behave remarkably different from chains, stars,
    branched or cross-linked polymers under pure, steady shear alone.
    They show a unique inflation in the vorticity direction because of a well-established backflow, which 
    highlights the importance of carefully realized hydrodynamics when dealing with such closed topologies.
    For rings with high contour length, an interplay between alignment at an angle to the flow axis and the backflow 
    causes the ring to self-stabilize in gradient direction and leads to a surcease of tumbling and a suppression of tank-treading.
    The entire object looks and behaves more like a rigid ring in this configuration and it behaves in this way even in the
    presence of complex knots along its backbone. On the basis of these phenomena, we suggest a potential 
    microfluidic filter to separate rings from chains, or rings of different sizes from each other. 
    As a knotted section on a sufficiently long ring has very sharp tank-treading-like transitions from one side of an open ring to another,
    we also envision that fluorescence techniques could be used to detect the presence of such a tight knot on a ring. 

    Although the present work was focused on single-molecule numerical experiments, future extensions should focus on the
    effects of shear and of the ring-stretching transition to the rheological behavior of dilute or semidilute ring polymer
    mixtures, as well as on the interplay between inflation and polymer rigidity. The former should be directly comparable
    with rheology experiments on ring polymer solutions, whereas the latter system can be realized by employing,
    e.g., short {DNA} rings.
    Work along these lines is currently under way.

    \subsection*{Acknowledgments}

    The authors thank Luca Tubiana for supplying them with his LocKnot program for knot localization,
        and Dimitris Vlassopoulos for a critical reading of the manuscript and helpful discussions.
        The computational results presented have been achieved in part using the Vienna Scientific Cluster (VSC). 
        M.L. has been supported in part by the uni:docs Doctoral Fellowship Programme of the University of Vienna.


    \footnotesize
    \bibliographystyle{unsrt}

\end{document}